\documentclass[preprintnumbers,article,amsmath,amssymb,floatfix,10pt,prd,twocolumn,
superscriptaddress,nofootinbib]{revtex4-2}
\usepackage{bm}
\usepackage{amsfonts}
\usepackage{latexsym}
\usepackage[latin1]{inputenc}
\usepackage{graphicx}
\usepackage{amsmath}
\usepackage{palatino}
\usepackage{mathpazo}
\usepackage{textcomp}
\usepackage{float}
\usepackage{booktabs}
\usepackage{bigints} 
\usepackage{dcolumn}
\usepackage{ragged2e}
\usepackage{hyperref}
\hypersetup{colorlinks,citecolor=blue}
\hypersetup{colorlinks=true,linkcolor=red,filecolor=magenta,    urlcolor=blue}
\usepackage{amsmath}
\usepackage{xcolor}
\usepackage{orcidlink}
\usepackage{epsfig}
\usepackage{subfigure}
\usepackage{commath}
\usepackage{mathtools}

\def\jnl@style{\it}
\def\aaref@jnl#1{{\jnl@style#1}}

\def\aaref@jnl#1{{\jnl@style#1}}

\def\aj{\aaref@jnl{AJ}}                   
\def\apj{\aaref@jnl{ApJ}}                 
\def\apjl{\aaref@jnl{ApJ}}                
\def\apjs{\aaref@jnl{ApJS}}               
\def\apss{\aaref@jnl{Ap\&SS}}             
\def\aap{\aaref@jnl{A\&A}}                
\def\aapr{\aaref@jnl{A\&A~Rev.}}          
\def\aaps{\aaref@jnl{A\&AS}}              
\def\mnras{\aaref@jnl{Mon.~Not.~Roy.~Astron.~Soc.}}             
\def\prd{\aaref@jnl{Phys.~Rev.~D}}        
\def\prc{\aaref@jnl{Phys.~Rev.~C}}  
\def\prl{\aaref@jnl{Phys.~Rev.~Lett.}}    
\def\qjras{\aaref@jnl{QJRAS}}             
\def\skytel{\aaref@jnl{S\&T}}             
\def\ssr{\aaref@jnl{Space~Sci.~Rev.}}     
\def\zap{\aaref@jnl{ZAp}}                 
\def\nat{\aaref@jnl{Nature}}              
\def\aplett{\aaref@jnl{Astrophys.~Lett.}} 
\def\apspr{\aaref@jnl{Astrophys.~Space~Phys.~Res.}} 
\def\physrep{\aaref@jnl{Phys.~Rep.}}      
\def\physscr{\aaref@jnl{Phys.~Scr}}       
\def\commat{\aaref@jnl{Comm.~Math.~Phys.}}              
\def\science{\aaref@jnl{Science}}               
\def\cqg{\aaref@jnl{Classical Quant.~Grav.}}            
\def\jpcs{\aaref@jnl{JPCS}}                                     
\def\ijmpd{\aaref@jnl{Int.~J.~Mod.~Phys.~D}}                    
\def\grg{\aaref@jnl{Gen.~Relat.~Gravit.}}               
\def\rpp{\aaref@jnl{Rep.~Prog.~Phys.}}          
\def\npa{\aaref@jnl{Nucl.~Phys.~A}}        
\def\lrr{\aaref@jnl{Living Rev.~Rel.}}                   
\def\jcap{\aaref@jnl{J.~Cosmology Astropart.~Phys.}}    
\def\rmp{\aaref@jnl{Rev.~Mod.~Phys.}}   
\def\epjc{\aaref@jnl{Eur.~Phys.~J.~C}} 
\def\plb{\aaref@jnl{~Phy.~Lett.~B}} 
\def\mpla{\aaref@jnl{Mod.~Phy.~Lett.~A}} 
\def\arxiv{\aaref@jnl{arxiv.org}}

\allowdisplaybreaks[1]

\begin{document}
\color{black}       
\title{A Study of stable wormhole solution with non-commutative geometry in the framework of linear $f(R,\mathcal{L}_m, T)$ gravity}

\author{Niklas Loewer}
\email{it99cabo@studserv.uni-leipzig.de}
\affiliation{Institut f\"ur Theoretische Physik, Universit\"at Leipzig,\\ Br\"uderstra\ss e 16, 04103 Leipzig, Germany.}

\author{Moreshwar Tayde\orcidlink{0000-0002-3110-3411}}
\email{moreshwartayde@gmail.com}
\affiliation{Department of Mathematics, Birla Institute of Technology and
Science-Pilani,\\ Hyderabad Campus, Hyderabad-500078, India.}

\author{P.K. Sahoo\orcidlink{0000-0003-2130-8832}}
\email{pksahoo@hyderabad.bits-pilani.ac.in}
\affiliation{Department of Mathematics, Birla Institute of Technology and
Science-Pilani,\\ Hyderabad Campus, Hyderabad-500078, India.}

%
\date{\today}
\begin{abstract}
This research delves into the potential existence of traversable wormholes (WHs) within the framework of modified, curvature based gravity. The modification includes linear perturbations of the matter Lagrangian and the trace of the energy-momentum tensor with specific coupling strengths $\alpha$ and $\beta$ and can thus be viewed as a special case of linear $f(R,T)$-gravity with a variable matter coupling or as the simplest additively separable $f(R,\mathcal{L}_m,T)$-model. A thorough examination of static WH solutions is undertaken using a constant redshift function; therefore, our work can be regarded as the first-order approximation of WH theories in $f(R,\mathcal{L}_m,T)$ . The analysis involves deriving WH shape functions based on non-commutative geometry, with a particular focus on Gaussian and Lorentzian matter distributions $\rho$. Constraints on the coupling parameters are developed so that the shape function satisfies both the flaring-out and asymptotic flatness conditions. Moreover, for positive coupling parameters, violating the null energy condition (NEC) at the WH throat $r_0$ demands the presence of exotic matter. For negative couplings, however, we find that exotic matter can be avoided by establishing the upper bound $\beta+\alpha/2<-\frac{1}{\rho r_0^2}-8\pi$. Additionally, the effects of gravitational lensing are explored, revealing the repulsive force of our modified gravity for large negative couplings. Lastly, the stability of the derived WH solutions is verified using the Tolman-Oppenheimer-Volkoff (TOV) formalism.
\end{abstract}

\maketitle


\section{Introduction}\label{ch: I}
\indent WHs represent solutions within gravitational theories that establish a connection between two distinct regions within a single spacetime or between two separate spacetimes via a throat (refer to \cite{I: VIS} for an extensive review). In the framework of General Relativity (GR), the formation of these WHs necessitates the presence of exotic matter \cite{I: MOR}, which inherently violates the null energy condition even at the classical level. It is worth noting that the total quantity of exotic matter required can be significantly minimized under specific conditions; see, for example, \cite{I: VIS1}. Various approaches exist to construct WHs, whether within General Relativity (as discussed in \cite{I: LEM}) or alternative gravitational theories. Examples of such theories include Brans-Dicke theory \cite{I: ANCHO}, the Randall-Sundrum model \cite{I: ANCHO1}, Braneworld \cite{I: CAMERA, I: BRONN, I: PARS, I: PARS1}, Born-Infeld theory \cite{I: RICH, I: SHAIKH}, Einstein-Cartan gravity \cite{I: BRONN1, I: MEHDI}, $f(R)$ gravity \cite{I: LOBO}, hybrid metric-Palatini gravity \cite{I: CAPO}, modified teleparallel gravity \cite{I: BOEH}, and $f(Q)$ gravity \cite{I: RASTGOO, I: BANERJEE, I: MUSTAFA, I: TAY3, I: GHOSH, I: TAY2}, among others. One of the benefits of working with modified theories of gravity is that the WH geometry can be supported by non-exotic matter, too \cite{I: MOTI}.\\
\indent Recent studies have delved into the topic of WHs within a more fundamental framework \cite{I: MALD, I: MALD1}, extending its relevance to condensed matter systems as well \cite{I: GONZ, I: ALEN}. Although direct observational evidence of WHs remains elusive, substantial progress has been made in their theoretical study. One intriguing line of inquiry is the potential existence of WHs within galactic halos, as explored in \cite{I: RAH, I: RAH1}. Research suggests that the observed flat rotation curves of galaxies could be consistent with a traversable WH geometry within the spacetime of a galactic halo. Further investigations have examined the possibility of detecting these WHs via gravitational lensing \cite{I: KUH}, with additional analyses presented in \cite{I: NAN}. Notably, supermassive black holes (BHs) at the centers of galaxies \cite{I: ERR, I: PAUL} are considered potential candidates for WHs, possibly originating from the early Universe.\\
\indent In $f(R)$ gravity, the traditional Einstein-Hilbert action is replaced by a function that depends arbitrarily on the Ricci scalar ($R$) \cite{I: SOTI, I: NOJIRI}. Falco et al. \cite{I: V. De Falco} constructed WH solutions in curvature-based extended theory, i.e., $f(R)$ gravity. Moreover, the authors \cite{I: V. De Falco1} examined static and spherically symmetric WH solutions within the frameworks of $f(R)$ metric, $f(T)$ teleparallel, and $f(Q)$ symmetric teleparallel models. Furthermore, $f(R, T)$ gravity emerges from coupling any function of the Ricci scalar $R$ and the trace $T$ of the corresponding stress-energy tensor with the matter Lagrangian density ($\mathcal{L}_m$) \cite{I: HARKO}. This theory is an extension of the $f(R)$ modified gravity framework \cite{I: HARKO}. The cosmological implications of $f(R, T)$ gravity have been extensively examined in the literature \cite{I: SHABANI, I: ZARE, I: MYRZA}. Additionally, black holes have been analyzed within the context of $f(R, T)$ gravity \cite{I: SANTOS}. Various studies have explored WH solutions within $f(R, T)$ gravity. For instance, Azizi \cite{I: AZIZI} derived a shape function based on the assumption of a linear equation of state (EoS) for matter, and the solutions provided in \cite{I: AZIZI} adhere to the energy conditions. Moraes and Sahoo \cite{I: MORAES} have also contributed to the modeling of WHs within the $f(R, T)$ gravity framework. Sharif and Nawazish conducted an in-depth analysis of spherically symmetric WH solutions, employing the Noether symmetry approach under the framework of $f(R, T)$ gravity \cite{I: SHARIF}. Further exploration of $f(R, T)$ gravity was undertaken in \cite{I: CHANDA}, where three different models were systematically investigated to derive exact WH solutions. The study presented in \cite{I: MORAES1} also focused on charged WHs within the extended $f(R, T)$ theory of gravity. Introducing charge into the WH framework adds complexity to the solutions, affecting both the stability and the overall structure of the WHs. Sharif and Fatima \cite{I: SHARIF1} advanced the field by introducing traversable WH solutions by applying the Karmarkar condition within $f(R, T)$ gravity. The Karmarkar condition, which is a criterion for embedding a four-dimensional spacetime in a higher-dimensional space, provides a novel method for constructing WH solutions. Moreover, Zubair et al. \cite{I: ZUBAIR} explored spherically symmetric wormhole solutions in modified $f(R,T)$ gravity by incorporating non-commutative geometry, expressed through Gaussian and Lorentzian string theory distributions. Additionally, the stability of the solutions was confirmed using an equilibrium condition.\\
\indent The $f(R)$ gravity framework has evolved to establish a direct relationship between the Ricci scalar $R$ and the matter Lagrangian $\mathcal{L}_m$, initially proposed by Bertolami et al. \cite{Ber}. Harko and Lobo further advanced this concept \cite{Har}, who introduced arbitrary matter-geometry couplings. These non-minimal curvature-matter couplings have been shown to have significant implications in both astrophysical and cosmological contexts \cite{Har1, Har4, Far}. Harko and Lobo \cite{Har5} further extended this idea by developing the $f(R, \mathcal{L}_m)$ modified gravity theory, which incorporates curvature-matter coupling into a broader theoretical framework. In this extended theory, the function $f(R, \mathcal{L}_m)$ is formulated as a general function that depends on both the Ricci scalar $R$ and the matter Lagrangian $\mathcal{L}_m$. A key consequence of this theory is that the energy-momentum tensor no longer has a zero covariant divergence. This non-zero divergence introduces an additional force component that acts orthogonally to the four-velocity vectors of the matter distribution, resulting in the deviation of a test particle's motion from its geodesic path. The introduction of $f(R, \mathcal{L}_m)$ gravity has generated significant interest within the scientific community, leading to numerous studies in cosmology and astrophysics \cite{Lab2, Har6, Sol, Lakhan}. These investigations have examined the various implications of this modified gravity framework, offering new opportunities for exploring the fundamental interactions between matter and spacetime.\\
\indent Recently, Haghani and Harko \cite{I: HAG} introduced a generalized theory of gravity that incorporates geometry-matter coupling, formulated as $f(R,\mathcal{L}_m, T)$. The proposed theory not only encompasses the $f(R)$, $f(R, T)$, and $f(R, \mathcal{L}_m)$ modified gravity theories as specific cases but also provides a broader framework for exploring these interactions. Building on this framework, Moraes et al. \cite{I: MEA} have derived WH solutions within the context of $f(R,\mathcal{L}_m, T)$ gravity and found that WHs can be filled with ordinary matter. 
The goal of our research is to investigate the physical implications of the proposed WH solutions using the simplest linear $f(R,\mathcal{L}_m,T)$ model that allows for additive decompositions $f(R,\mathcal{L}_m,T)=g(R)+h(\mathcal{L}_m)+i(T)$, which can be viewed as a first order approximation of the general theory serving as a bridge between $f(R,T)$ and $f(R,\mathcal{L}_m)$ gravity. A nice feature of this approach is the existence of analytical solutions when assuming a constant redshift function. In this way, this paper can serve as a starting point for further studies of WHs in $f(R,\mathcal{L}_m,T)$ gravity.
 Given the relatively limited exploration of this gravity theory within the astrophysical domain, we are motivated to contribute to the field by investigating WH solutions supported by non-commutative geometry within the framework of $f(R,\mathcal{L}_m, T)$ gravity. This study aims to expand our understanding of how non-commutative geometry can influence the existence and characteristics of WHs in this modified gravity context.\\
\indent In recent years, non-commutative geometry has garnered considerable interest from the scientific community, establishing itself as a crucial element in the study of space-time geometry, with far-reaching consequences across multiple fields. This branch of mathematics provides a robust framework for analyzing the characteristics of compact objects, such as WHs. The Parikh-Wilczek tunneling mechanism in the context of non-commutative higher-dimensional black holes has been thoroughly investigated in Ref. \cite{I: NOZA}. Sushkov explored WHs sustained by phantom energy, utilizing a Gaussian distribution, as detailed in Ref. \cite{I: SUSH}. Rahaman et al. \cite{I: RAH2} delved into WH solutions within a Gaussian distribution setting, demonstrating their existence exclusively in four and five-dimensional spacetimes. Further, WH solutions have been evaluated under non-commutative geometries in $f(T)$ \cite{I: RANI} and $f(Q)$ gravity theories \cite{I: HAS}. Additionally, the BTZ black hole within a non-commutative framework was the subject of study in Ref. \cite{I: RAH3}. Moreover, Tayde et al. \cite{I: TAY} explored WH solutions within the context of non-commutative geometry for both linear and non-linear forms of $f(Q, T)$ gravity. They conducted an analytical investigation for the linear model and employed numerical methods for the non-linear case. Additionally, they studied the gravitational lensing effect associated with the exact WH solution, finding that the deflection angle becomes infinite at the WH throat. In addition, WHs with non-commutative geometry and conformal symmetry within the framework of $f(R,\mathcal{L}_m)$ gravity were examined in Ref. \cite{I: KAVYA, I: KAVYA3}. For readers interested in delving deeper into this topic, further insights and detailed discussions can be found in the references provided in \cite{I: TAY1, I: KAVYA1, I: KAVYA2}. These resources offer comprehensive analyses and additional context for those seeking a more thorough understanding of the subject matter.\\
\indent The organization of this paper is as follows: Section \ref{ch: II} outlines the criteria for defining a traversable WH and introduces the formalism of $f(R,\mathcal{L}_m, T)$ gravity, focusing on deriving the field equations within a linear model framework. Section \ref{ch: III} investigates non-commutative matter distributions, particularly of Gaussian and Lorentzian profiles, and investigates the traversability condition of the corresponding solutions to the field equations. Section \ref{ch: IV} examines the energy conditions associated with both distributions. In order to demonstrate the physical implication of our theory, Section \ref{ch: V} extends the analysis to include the effects of gravitational lensing. Section \ref{ch: VI} presents a stability analysis using the TOV equation. Finally, Section \ref{ch: VII} concludes with a summary of our findings and a detailed discussion of the implications of the results obtained from this study.

\section{Field Equations for spherically symmetric wormholes in $f(R,\mathcal{L}_m,T)$ gravity}
\label{ch: II}

The most general static, spherically symmetric spacetime is given by the following metric \cite{II: CAR}:
\begin{equation} \label{1}
ds^{2}=-e^{\nu(r)}dt^{2}+e^{\lambda(r)} dr^{2}+r^{2} d\theta^{2}+r^{2}\sin^{2}\theta d\phi^{2},
\end{equation}
where $\left\{t\in\mathbb{R},r\in[r_0,\infty),\theta\in(0,\pi),\phi\in [0,2\pi)\right\}$ and both $\nu(r)$ and $\lambda(r)$ and smooth functions of the radial coordinate. For our discussion, we will work with $\nu(r)=2\Phi(r)$, where $\Phi(r)$ is the \textit{redshift} function. The terminology originates from the fact that when calculating the gravitational redshift for the metric given by Eq. \eqref{1}, $g_{tt}$ is used to define the energy. Thus, $\Phi$ exactly characterizes the redshift. Furthermore, we make use of the (within WH literature) common choice of $e^{\lambda(r)}=(1-b(r)/r)^{-1}$, stemming from \cite{I: MOR}. Here, $b(r)$ is called the \textit{shape-function}, rendering a traversable WH if the following conditions are satisfied \cite{I: MOR, II: HAS}: 
\begin{equation}
    \begin{split}
        \frac{b(r)}{r}\!<\!1\,\,\forall r\!>\!r_0;\,b(r_0)\!=\!r_0\quad&\textit{Throat Condition}\\
        \frac{b-b'r}{b^2}\biggr\rvert_{r_0}\!\!>\!0\rightarrow b'(r_0)\!<\!1, \quad&\textit{Flaring-Out Condition}\\
        \lim_{r\rightarrow\infty}\frac{b(r)}{r}=0\quad&\textit{Asymptotic Flatness Condition}.\\
    \end{split}
    \label{eq: TWC}
\end{equation}

Additionally, the redshift function must be finite for all $r$ (preventing the formation of horizons), which will be trivially met by our choice of $\Phi(r)$ in the later discussion. Naturally, the metric given by Eq. \eqref{1} must also solve the Einstein Equations; therefore, given a suitable energy-momentum-tensor, one obtains the functions $b(r)$ and $\Phi$. However, as Morris and Thorne \cite{I: MOR} already pointed out, traversable WH solutions can only be obtained when allowing so-called \textit{exotic} matter, i.e., a matter that violated various energy conditions. Now, this problem can be circumvented by modifying the underlying Einstein Equations. When working with curvature as the defining quantity, this approach is known as $f(R)$ gravity; in this case, though, either the wormholes cannot be asymptotically flat or the effective gravitational constant must change its sign somewhere in some region of spacetime \cite{I: BSS}. The latter restriction is strongly reflected in our examination of the lensing effects, see Section \ref{ch: V}. Our model further wishes to employ a coupling to trace-term $T$ for the energy-momentum-tensor as well as a linear modification to the matter Lagrangian $\mathcal{L}_m$; technically, the resulting model belongs to the class of renormalized $f(R,T)$, since the standard matter coupling constant is replaced by an arbitrary one. However, because our equations are derived within the $f(R,\mathcal{L}_m, T)$ setting tendered by Haghani and Harko \cite{I: HAG} and intend to serve as a linear approximation of a more refined $f(R,\mathcal{L}_m, T)$ theory, we will refer to our gravitational setting as $f(R,\mathcal{L}_m, T)$ gravity.

Now, to begin with, we propose the gravitational action functional \cite{I: HAG}:
\begin{equation}\label{eq: Ac}
\mathcal{S}=\frac{1}{16\pi}\int\,f(R,\mathcal{L}_m,T)\sqrt{-g}\,d^4x+\int \mathcal{L}_m\,\sqrt{-g}\,d^4x\,.
\end{equation}
In units where $G=c=1$. Here, $R$ represents the Ricci scalar - the curvature quantity - which is linked to the metric tensor $g_{\mu\nu}$ with determinant $g$. To explicitly see the connection between $R$ and $g_{\mu\nu}$, consider the torsion-less, metric-affine connection
\begin{equation}
    \Gamma^\alpha_{\beta\gamma}= \frac{1}{2} g^{\alpha\lambda} \left( \frac{\partial g_{\gamma\lambda}}{\partial x^\beta} + \frac{\partial g_{\lambda\beta}}{\partial x^\gamma} - \frac{\partial g_{\beta\gamma}}{\partial x^\lambda} \right).
    \label{eq: CHR}
\end{equation}
This connection is used to define a covariant derivative and, in this way, encapsulates the properties of the curved geometry. Constructing the \textit{Ricci}-tensor
\begin{equation}\label{eq: RIC}
R_{\mu\nu}= \partial_\lambda \Gamma^\lambda_{\mu\nu} - \partial_\nu \Gamma^\lambda_{\lambda\mu} + \Gamma^\sigma_{\mu\nu} \Gamma^\lambda_{\sigma\lambda} - \Gamma^\lambda_{\nu\sigma} \Gamma^\sigma_{\mu\lambda}\,,
\end{equation} its contraction, the Ricci scalar
\begin{equation}\label{eq: RICS}
R=g^{\mu\nu} R_{\mu\nu}\,,
\end{equation}
supplies us with a coordinate-independent curvature quantity. Thus, we obtain an action functional that is manifestly invariant under the symmetry transformations of spacetime (diffeomorphism) if the matter Lagrangian is also coordinate independent. The resulting equations of motion, our modified Einstein equations, are obtained by varying $\mathcal{S}$ w.r.t. $g_{\mu\nu}$ and demanding a stationary action:
\begin{multline}\label{6}
\hspace{-0.5cm}f_{R}R_{\mu \nu }-\frac{1}{2}\left[ f-(f_{\mathcal{L}_m}+2f_{T})\mathcal{L}_m\right] g_{\mu \nu } +\left( g_{\mu \nu }\Box -\nabla _{\mu }\nabla _{\nu }\right) f_{R}\\
=\left[ 8\pi +\frac{1}{2}(f_{\mathcal{L}_m}+2f_{T})\right] T_{\mu \nu } +f_{T}\tau _{\mu \nu }, 
\end{multline}
where, $f_{R}=\frac{\partial f(R,\mathcal{L}_m,T)}{\partial R}$, $f_{\mathcal{L}_m}=\frac{\partial f(R,\mathcal{L}_m,T)}{\partial \mathcal{L}_m}$, $f_{T}=\frac{\partial f(R,\mathcal{L}_m,T)}{\partial T}$ and $\tau _{\mu \nu }=2g^{\alpha \beta }\frac{\partial ^{2}\mathcal{L}_m}{\partial g^{\mu \nu }\partial g^{\alpha \beta }}$. \\
Additionally, $T_{\mu\nu}$ denotes the energy-momentum tensor for the cosmic fluid, and it is given by:
\begin{equation}
T_{\mu\nu} = \frac{-2}{\sqrt{-g}} \frac{\delta(\sqrt{-g}\mathcal{L}_m)}{\delta g^{\mu\nu}}.
\end{equation}\\
We assume this fluid to be anisotropic and of the form
\begin{equation}
\label{EMT}
T^\mu_{\,\,\,\,\nu} =\text{diag}(-\rho,p_r,p_t,p_t),
\end{equation}
so that (following \cite{I: MOR, I: VIS, I: KUH})
\begin{equation}
T_{\mu\nu}=\left(\rho+p_t\right)u_{\mu}\,u_{\nu}+p_t\,\delta_{\mu\nu}+\left(p_r-p_t\right)v_{\mu}\,v_{\nu}\,,
\end{equation}
where $u_\mu$ is the four-velocity and $v_\mu$ is a unit spacelike covector, both normalized to $\mp1$, respectively. Furthermore, $\rho$ corresponds to the energy density, and the radial and tangential pressures $p_r$ and $p_t$ are functions of $r$. \\
Then, the modified Einstein equations \eqref{6} read:
\begin{multline}\label{11}
\left( 1-\frac{b}{r} \right) \left[ \left\lbrace  \Phi''+\Phi'^2   - \frac{(rb'-b)}{2r(r-b)}\Phi' + \frac{2\Phi'}{r}\right\rbrace f_R
\right.\\ \left.
-\left\lbrace \Phi' - \frac{(rb'-b)}{2r(r-b)}  +\frac{2}{r}\right\rbrace f_R' - f_R'' \right] + \frac{1}{2} \left( f- f_{\mathcal{L}_m} {\mathcal{L}_m} \right) \\
= \frac{1}{2} \rho \left(16 \pi +  f_{\mathcal{L}_m}+2f_{T}\right)+ f_{T} \mathcal{L}_m\,,
\end{multline}
\begin{multline}\label{12}
\left( 1-\frac{b}{r} \right) \left[ \left\lbrace  - \left(\Phi''+\Phi'^2\right)  + \frac{(rb'-b)}{2r(r-b)} \left( \Phi' +\frac{2}{r} \right) \right\rbrace f_R  
\right.\\ \left.
+ \left\lbrace \Phi'- \frac{(rb'-b)}{2r(r-b)}  +\frac{2}{r}\right\rbrace f_R' \right] - \frac{1}{2} \left( f- f_{{\mathcal{L}_m}} {\mathcal{L}_m} \right) \\
= \frac{1}{2} p_r \left(16\pi+f_{\mathcal{L}_m}+2f_{T}\right)- f_{T} \mathcal{L}_m\,,
\end{multline}
\begin{multline}\label{13}
\left( 1-\frac{b}{r} \right) \left[ \left\lbrace - \frac{\Phi'}{r}  + \frac{(rb'+b)}{2r^2(r-b)} \right\rbrace f_R+ \left\lbrace \Phi'+\frac{2}{r} \right.\right.\\ \left.\left.
- \frac{(rb'-b)}{2r(r-b)}  \right\rbrace f_R' + f_R'' \right] - \frac{1}{2} \left( f-{\mathcal{L}_m} f_{{\mathcal{L}_m}} \right) \\
= \frac{1}{2} p_t \left(16\pi+f_{\mathcal{L}_m}+2f_{T}\right)- f_{T} \mathcal{L}_m\,.
\end{multline}
To study these equations, we use the following linear model:
\begin{equation}\label{14}
f(R,\mathcal{L}_m,T) = R + \alpha \mathcal{L}_m + \beta T,
\end{equation} 
where the parameters $\alpha$ and $\beta$ determine the strength of our modification. It is easy to see from Eq. \eqref{eq: Ac} that for $\alpha=\beta=0$, our model reduces to regular general relativity.
In order to receive analytical solutions to the Equations \eqref{11}-\eqref{13}, we assume a constant redshift $\Phi(r)\equiv \Phi$. Solutions for traversable WHs with constant redshift have previously been obtained for $f(R)$ gravity \cite{II: GOD, II: GOD2} and $f(R,\mathcal{L}_m)$ gravity \cite{Sol, Lakhan}. Finally, calculating the Ricci scalar for the metric \eqref{1},
\begin{multline} \label{15}
R=\frac{2b'}{r^2} - 2 \left\lbrace \Phi''+\Phi'^2+\frac{\Phi'}{r} \right\rbrace \left( 1-\frac{b}{r} \right) \\
+ \frac{\Phi'}{r^2} \left( rb'+b-2r \right),
\end{multline}
and setting $\mathcal{L}_m=-\rho$ (following Haghani \cite{I: HAG}), we find the relations:
\begin{equation}\label{14a}
    \frac{b'}{r^2}=\rho  \left(\lambda -\frac{\beta }{2}\right)-\frac{\beta  }{2} (p_r+2 p_t)\,,
\end{equation}
\begin{equation}\label{14b}
    -\frac{b}{r^3}=p_r \left(\lambda + \frac{\beta }{2} \right)+\frac{ \beta}{2}  (2 p_t+\rho )\,,
\end{equation}
\begin{equation}\label{14c}
   \frac{1}{2 r^2}\left(\frac{b}{r}-b'\right)=p_t ( \lambda+\beta  ) +\frac{ \beta}{2}  (p_r+\rho ) \,,
\end{equation}
where $\lambda=8 \pi +\frac{\alpha }{2}+\beta $.
The results match the one given by Moraes et al. \cite{I: MEA} up to a change of sign of the second term in the first equation. Now, it is straightforward to find $\rho, p_r$ and $p_t$ in terms of $b$; e.g. by adding the Eq. \eqref{14a} and \eqref{14b}, one obtains an expression $(\rho+p_r)$, which can then be used to find $p_t$ via Eq. \eqref{14c}. In that way, the system of Equations \eqref{14a}-\eqref{14c} simplifies to:
\begin{equation}\label{15a}
    \rho=\frac{b'}{\lambda r^2}\,,
\end{equation}
\begin{equation}\label{15b}
    p_r=-\frac{b}{\lambda r^3}\,,
\end{equation}
\begin{equation}\label{15c}
   p_t=\frac{b-b'r}{2\lambda r^3} \,.
\end{equation} Comparing this result to Moraes et al. \cite{I: MEA}, the equality of the final expressions in both works solidifies the answer, further suggesting that the disagreement of Eq. \eqref{14a} arose merely from a typographical error in their work, as we respectfully note.

\section{Specific solutions for non-commutative matter distributions}
\label{ch: III}

\begin{figure*}[t]
    \centering
    \includegraphics[width=13cm,height=5cm]{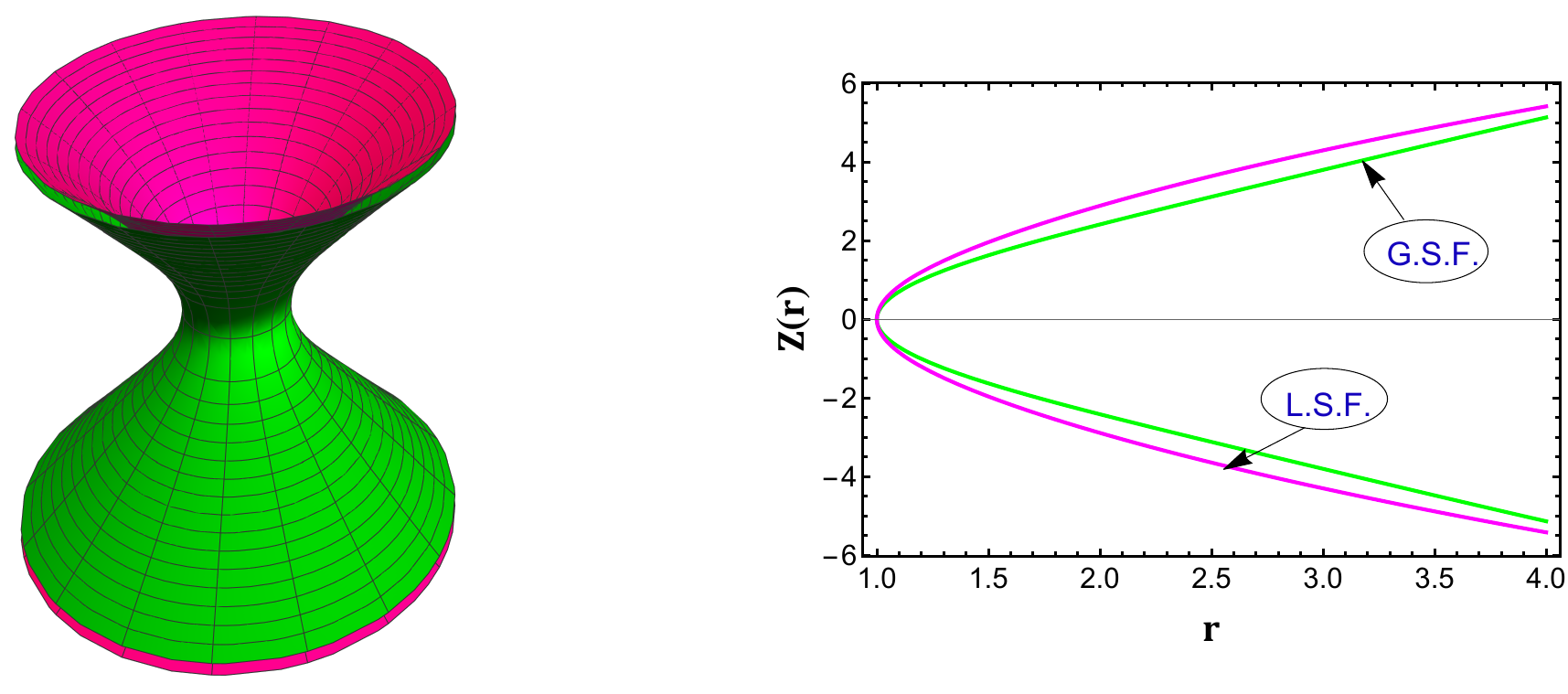}
    \caption{The figure displays the embedding diagram for both distributions. Additionally, the parameters, specifically $\alpha = 1$, $\beta=1$, $\Theta = 2$, $M = 1$, and $r_0 = 1$, are held constant.}
    \label{fig: E}
\end{figure*}

Motivated by the apparent formulation of black holes at the Planck scale \cite{III: DFR}, one can propose a quantization of the spacetime coordinates, realized by postulating the commutation relations
\begin{equation}
\label{C}
    [x^\mu,x^\nu]=i\Theta^{\mu\nu},
\end{equation}
where $\Theta^{\mu\nu}$ is of order $\ell_p^2$ ($\ell_p$ is the Planck length). Via the associated uncertainty relation, this confines spacetime to discrete volume elements. Discretization of spacetime is an intrinsic feature of many of the prominent candidates for a theory of quantum gravity, such as String Theory \cite{III: SCHOM, III: CHONG, III: SEIBE}, Loop Quantum Gravity \cite{III: AMELI} and quantum field theory in curved spacetime \cite{III: SZA, III: BOR, III: MUC}. It has been shown by Smailagic and Spalluci in \cite{III: SMA} that the Commutation Relations \eqref{C} transform point-like mass distributions into Gaussians of the form
 \begin{equation}\label{21}
 \rho(r) =\frac{M e^{-\frac{r^2}{4 \Theta }}}{\left(4\pi\Theta\right)^{3/2}}\,,
 \end{equation}
 where $\sqrt{\Theta}\sim\ell_p$ is the non-commutativity parameter and $M$ is the total mass of the particle. Hence, to incorporate possible quantum features of the WH, we will investigate the WH geometry using this Gaussian mass distribution. This approach has also been applied for black holes \cite{III:  NIC}, $f(T)$ \cite{I: RANI}, $f(Q)$ \cite{I: HAS}, $f(Q,T)$ \cite{I: TAY}, $f(R,T)$ \cite{I: ZUBAIR}, and $f(R,\mathcal{L}_m)$ gravity \cite{I: KAVYA}. Surprisingly, in a similar study of the entropic force of Schwarzschild black holes \cite{III: MEH}, the author notes that when smearing the particle with a Lorentzian mass distribution
  \begin{equation}\label{22}
 \rho(r)=\frac{\sqrt{\Theta } M}{\pi ^2 \left(\Theta +r^2\right)^2}\,
 \end{equation}
 instead of the Gaussian, he receives physically equivalent results. Therefore, we, too, will apply our calculations to Lorentzian mass distribution as a means to compare results within our theory. \\
 In Figure \ref{fig: E}, a spacetime-embedding diagram is shown for both the Gaussian and Lorentzian mass distributions, sketching the geometrical appearance of the WH. The embedding surface $Z(r)$ is defined by \cite{I: MOR}:
 \begin{equation}
      Z(r)=\pm\bigintss_{\,r_0}^{\infty}\biggl(\frac{r}{b(r)}-1\biggr)^{-\frac{1}{2}}\text{d}r,
 \end{equation}
where the different signs denote the two branches of spacetime connected by the WH.

\subsection{Gaussian distribution}

\begin{figure*}[t]
\begin{tabular}{ccc}
\subfigure[$b(r)$]{{\includegraphics[width=0.3\textwidth]{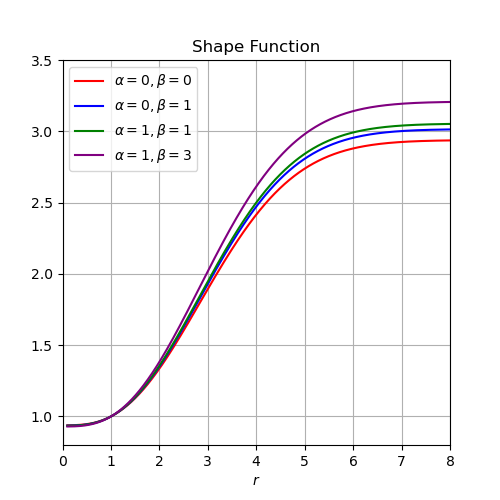}}
\label{fig: SF}}&
\subfigure[$b'(r)$]{{\includegraphics[width=0.3\textwidth]{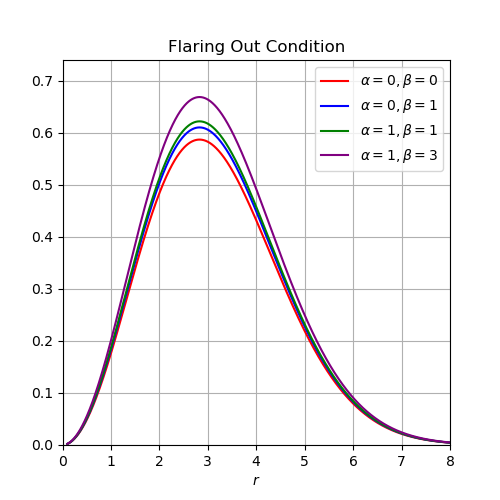}}
\label{fig: FOC}}&
\subfigure[$b(r)/r$]{{\includegraphics[width=0.3\textwidth]{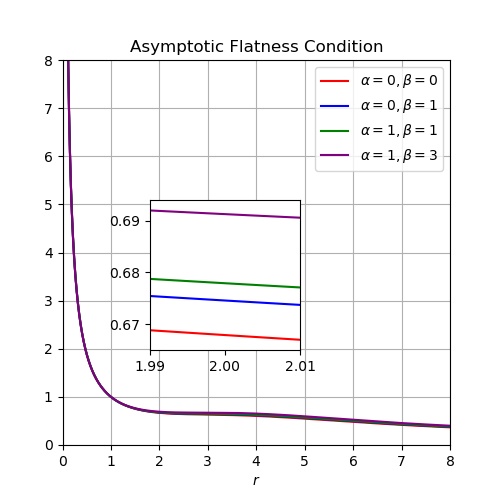}}
\label{fig: AFC}}
\end{tabular}
\caption{The figure illustrates the conditions of the shape function relative to the radial coordinate $r$ for different values of $\alpha$ and $\beta$ [showing GR $\left(\alpha=0, \beta=0\right)$, $f(R,T)$ $\left(\alpha=0, \beta=1\right)$, and $f(R,\mathcal{L}_m,T)$ $\left(\alpha=1, \beta\in\lbrace{1,3\rbrace}\right)$ cases] under the Gaussian distribution. The other parameters, specifically $\Theta = 2$, $M = 1$, and $r_0 = 1$, are held constant.}
\label{fig: C}
\end{figure*}

We start by comparing Eq. \eqref{15a} to our choice of mass distribution, Eq. \eqref{21}:
 \begin{equation}\label{23}
 \frac{b'(r)}{\lambda r^2}=\frac{M e^{-\frac{r^2}{4 \Theta }}}{\left(4\pi\Theta\right)^{3/2}}\,.
\end{equation}
Now, integrating the shape function and eliminating the integration constant by imposing the throat condition $b(r_0)=r_0$, we find:
\begin{equation}
\label{b}
\hspace{-0.4cm}\textbf{(G.S.F.)}\,\,\,\, b(r)=r_0+\frac{\lambda M}{4\pi}\biggl(\text{erf}\Bigl(\frac{r}{2 \sqrt{\Theta }}\Bigr)-\frac{re^{-\frac{r^2}{4 \Theta }}}{\sqrt{\pi\Theta}}+\mathcal{K}_G\biggr),
\end{equation}
where $\mathcal{K}_G=\frac{r_0e^{-\frac{r_0^2}{4\Theta}}}{\sqrt{\pi\Theta}}-\text{erf}\left(\frac{r_0}{2 \sqrt{\Theta }}\right)$ and ``erf" is the Gaussian error function defined by 
\begin{equation}
\label{erf}
    \text{erf}(x)=\frac{2}{\sqrt{\pi}}\int\limits_{0}^{x}e^{-t^2}dt.
\end{equation}
In Figure \ref{fig: SF}, the shape function is plotted for different values of the trace coupling $\beta$; the other parameters are kept constant.
On the contrary, the first derivative of the shape function can  trivially be obtained by rearranging Eq. \eqref{23}:
\begin{equation}
    b'(r)=\frac{\lambda M r^2 e^{-\frac{r^2}{4 \Theta }}}{\left(4\pi\Theta\right)^{3/2}}.
\end{equation}
Let us quickly check the Traversability Conditions \eqref{eq: TWC} for the Gaussian shape function. On the one hand, the Flaring-Out Condition (FOC) reads
\begin{equation}
    \frac{\lambda M}{(4\pi\Theta)^{3/2}}r_0^2e^{-\frac{r_0^2}{4\Theta}}<1,
\end{equation}
or, in terms of the first order coupling parameters
\begin{eqnarray}
    \beta+\frac{\alpha}{2}<\frac{(4\pi\Theta)^{3/2}}{Mr_0^2}e^{\frac{r_0^2}{4\Theta}}-8\pi.
\end{eqnarray}
Clearly, in the non-commutativity limit $\Theta\rightarrow0$, all trace couplings will satisfy the FOC. Setting $r_0=M=1$ and $\Theta=2$, as shown in Figure \ref{fig: FOC}, one needs $\beta+\frac{\alpha}{2}\gtrapprox118$ to violate the FOC. Asymptotic flatness, on the other hand, is always satisfied, as can be verified by taking the limit of $b(r)/r$ as $r\rightarrow\infty$. By the definition of the error function, Eq. \eqref{erf}, this term vanishes in the limit, as well as the exponential terms and the constants. This behavior is also visible in Figure \ref{fig: AFC}, where once again $r_0=M=1$ and $\Theta=2$. However, it is unclear whether $b(r)/r<1$ should be satisfied everywhere. Analytically, this relation cannot be solved; however, the slight maximum of the graph in Figure \ref{fig: AFC} suggests the existence of a set of parameters s.t. $b(r)/r\geq 1$. And indeed, for $r_0=M=1$ and $\Theta=2$, a value of $\beta\gtrapprox28$ causes a violation of the condition $b(r)/r<1$ for $\alpha=1$. The non-commutativity limit recovers the correct behavior, as only the $r_0$ term in Eq. \eqref{b} survives.

\subsection{Lorentzian distribution}

\begin{figure*}[t]
\begin{tabular}{ccc}
\subfigure[$b(r)$]{{\includegraphics[width=0.3\textwidth]{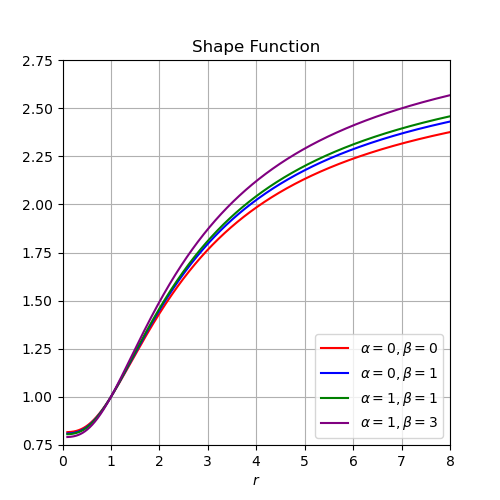}}
\label{fig: SFL}}&
\subfigure[$b'(r)$]{{\includegraphics[width=0.3\textwidth]{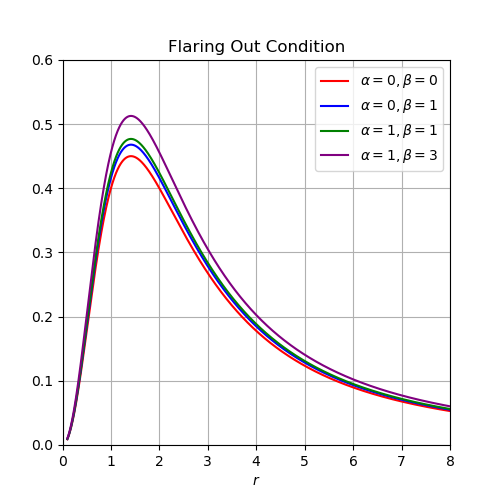}}
\label{fig: FOCL}}&
\subfigure[$b(r)/r$]{{\includegraphics[width=0.3\textwidth]{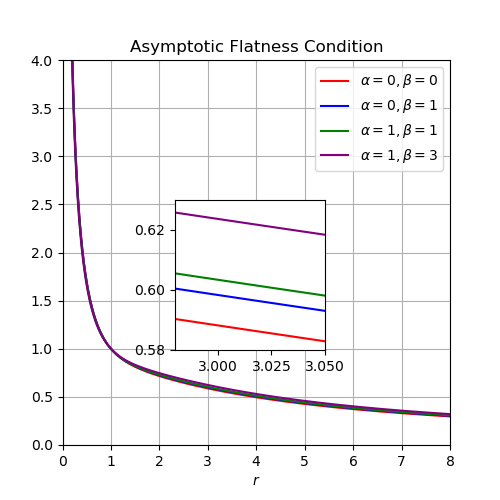}}
\label{fig: AFCL}}
\end{tabular}
\caption{The figure illustrates the conditions of the shape function relative to the radial coordinate $r$ for different values of $\alpha$ and $\beta$ [showing GR $\left(\alpha=0, \beta=0\right)$, $f(R,T)$ $\left(\alpha=0, \beta=1\right)$, and $f(R,\mathcal{L}_m,T)$ $\left(\alpha=1, \beta\in\lbrace{1,3\rbrace}\right)$ cases] under the Lorentzian distribution. The other parameters, specifically $\Theta = 2$, $M = 1$, and $r_0 = 1$, are held constant.}
\label{fig: CL}
\end{figure*}

In this subsection,  we will discuss the WH solution under non-commutative geometry using the Lorentzian distribution given by Eq. \eqref{22}. 
Similarly to the previous chapter, we obtain the differential equation
 \begin{equation}\label{33}
  \frac{b'(r)}{\lambda r^2}=\frac{\sqrt{\Theta } M}{\pi ^2 \left(\Theta +r^2\right)^2}\,.
 \end{equation}
 Integrating the above equation with the throat condition $b(r_0)=r_0$ yields
 \begin{equation}
 \label{34}
 \hspace{-0.4cm}\textbf{(L.S.F)}\,\,\,\, b(r)=r_0+\frac{\lambda M}{2\pi^2} \biggl(\tan ^{-1}\Bigl(\frac{r}{\sqrt{\Theta }}\Bigr)-\frac{\sqrt\Theta r}{\Theta+r^2}+\mathcal{K}_L\biggr),
 \end{equation}
where $\mathcal{K}_L=\frac{\sqrt\Theta r_0}{\Theta+r_0^2}-\tan ^{-1}\left(\frac{r_0}{\sqrt{\Theta }}\right)$. Figure \ref{fig: SFL} displays the function for the same set of parameters used before. Here, especially the large $r$ behaviour differs from the Gaussian shape function. Again, $b'(r)$ can easily be read off Eq. \eqref{33}:
\begin{equation}
    b'(r)=\frac{\lambda \sqrt{\Theta} M r^2}{\pi^2(\Theta+r^2)^2}.
\end{equation}
Therefore, the FOC will be satisfied for all
\begin{equation*}
    \beta+\frac{\alpha}{2}<\frac{\pi^2(\Theta+r_0^2)^2}{\sqrt{\Theta}Mr_0^2}-8\pi,
\end{equation*}
amounting to $\beta+\frac{\alpha}{2}\lessapprox38$ for our standard choice of parameters introduced in the previous subsection. The non-commutativity limit trivializes this relation for all finite trace and matter couplings. In Figure \ref{fig: FOCL}, the FOC is plotted for different combinations of $\alpha$ and $\beta$, next to a graph of $b(r)/r$ in Figure \ref{fig: AFCL}. Since the limit of $b(r)/r$ is zero as $r\rightarrow\infty$, the Lorentzian Morris-Thorne WH is asymptotically flat for all possible couplings, as the figure indicates. One can easily verify that the non-commutativity limit does not interfere with this result. However, problems arise again when demanding $b(r)/r<1$ everywhere because there is no analytical solution to this relation. Hence, we must use numerical methods and find that, e.g., regarding the parameters of Figure \ref{fig: AFCL} and taking $\alpha=1$, $\beta\gtrapprox32$ s.t. $b(r)/r\geq1$ for some $r>r_0$. In the limit $\Theta\rightarrow0$, the terms in the bracket of Eq. \eqref{34} cancel, and $b(r)/r<1$ is trivially satisfied for all $r>r_0$.

\section{Energy Conditions}
\label{ch: IV}
Within the framework of General Relativity, energy conditions serve as generalizations of the physical statement that energy should be positive in terms of the energy-momentum tensor $T_{\mu\nu}$. There are indeed various generalizations of different degrees of universality. For the anisotropic fluid we assumed in Eq. \eqref{EMT}, the relevant energy conditions read \cite{I: TAY,I: TAY1}:
\begin{multline}
        \textbf{Null Energy Condition:}\shoveleft{}\\
        \shoveright{\rho+p_r\geq0,\quad\rho+p_t\geq0.}\\
        \shoveleft{\textbf{Weak Energy Condition:}}\\       \shoveright{\rho\geq0,\quad\rho+p_r\geq0,\quad\rho+p_t\geq0.}\\
        \shoveleft{\textbf{Dominant Energy Condition:}}\\
        \shoveright{\rho\geq0,\quad\rho-|p_r|\geq0,\quad\rho-|p_t|\geq0.}\\
        \shoveleft{\textbf{Strong Energy Condition:}}\\
\shoveright{}\rho+p_r+2p_t\geq0,\quad\rho+p_r\geq0,\quad\rho+p_t\geq0.
\end{multline}
For a derivation and review of these conditions in $f(R)$ gravity, see \cite{IV: CAP}. In modified gravity, the energy conditions can be geometrically retrieved via the Raychaudhuri equation \cite{IV: RAY} for an effective stress-energy tensor. For WHs, the significant constraint is given by the Null Energy Condition (NEC); a violation of this at the throat implies the existence of exotic matter \cite{IV: HAR}. Recalling the expressions for $\rho$, $p_r$ and $p_t$ given by Eq. \eqref{14a}-\eqref{14c}, the NEC (at $r=r_0$) can be rewritten as:
\begin{equation}
\begin{split}
\label{NECr}
        \frac{b'(r_0)}{\lambda r_0^2}-\frac{1}{\lambda r_0^2}\geq0\\
        \frac{b'(r_0)}{2\lambda r_0^2}+\frac{1}{2\lambda r_0^2}\geq0
        \end{split}
\end{equation}
However, assuming positive $\lambda$, the first statement contradicts the Flaring-Out Condition (while the second becomes trivial). Hence, the only way we can exclude exotic matter from our theory of traversable WHs is by assuming $\lambda<0$, i.e.:
\begin{equation}
    \beta+\alpha/2<-8\pi.
\end{equation}
In this fashion, the first line of Eq. \eqref{NECr} is just the statement that FOC must be satisfied (see Ch. \ref{ch: III} for the respective constraints arising from that). Furthermore, the second line demands
\begin{equation}
    \begin{split}
    \label{eq: NEC}
        &\rho\geq-\frac{1}{\lambda r_0^2}\\
        \longleftrightarrow \quad& \lambda<-\frac{1}{\rho r_0^2}.
    \end{split}
\end{equation}
This result is in accordance with the literature \cite{I: MEA, IV: ROS}. It is a common approach to denote the term $\lambda\rho=b'/r^2$ as an effective mass density $\rho_{eff}$ (and similarly for the pressure components) s.t. the NEC/FOC constraint is equivalent to the statement that there exists an effective energy-momentum tensor, resulting from the non-standard couplings in Eq. \eqref{eq: Ac}, for which the NEC is violated, as well as a classical energy-momentum tensor which satisfies the NEC. In simple terms, one could say that the WH geometry is supported not by the regular matter but by its non-standard gravitational interaction.\\
Lastly, let us briefly examine the other energy conditions. An immediate conclusion from the above statements is that the Weak Energy Condition must be satisfied whenever the NEC is since $\rho$ is chosen to be positive! Furthermore, from Eq. \eqref{15a}-\eqref{15c} we see that
\begin{equation}
\label{eq:SEC}
    \rho+p_r+2p_t=0\,.
\end{equation}
Hence, the strong energy condition depends only on the NEC. To see whether the same result can be achieved for the Dominant Energy Conditions (DEC), write
\begin{equation*}
\begin{split}
    \rho-|p_r|\bigg\vert_{r=r_0}&=\frac{b'(r_0)}{\lambda r_0^2}-\abs{\frac{1}{\lambda r_0^2}}\\
     \rho-|p_t|\bigg\vert_{r=r_0}&=\frac{b'(r_0)}{\lambda r_0^2}-\abs{\frac{1}{2\lambda r_0^2}-\frac{b'(r_0)}{2\lambda r_0^2}}.
\end{split}
\end{equation*}
Recalling that the FOC and the first line of the NEC, Eq. \eqref{NECr}, demand $\lambda<0$, the radial DEC reads:
\begin{equation}
    \frac{b'(r_0)}{\lambda r_0^2}+\frac{1}{\lambda r_0^2}\geq0,
\end{equation}
which, of course, is just the tangential NEC. Additionally, from the radial NEC, it follows that 
\begin{equation*}
    \frac{1}{2\lambda r_0^2}-\frac{b'(r_0)}{2\lambda r_0^2}<0
\end{equation*}
and thus,
\begin{eqnarray}
    \rho-|p_t|\bigg\vert_{r=r_0}=\frac{b'(r_0)}{2\lambda r_0^2}+\frac{1}{2\lambda r_0^2},
\end{eqnarray}
being greater or equal to zero if and only if the tangential NEC is satisfied, too. Hence, for our mass distributions, all energy conditions strictly follow the NEC; this is another reason why, in the upcoming subsections, we will only numerically study the NEC.

\subsection{Gaussian Distribution}

\begin{figure*}[t]
\begin{tabular}{ccc}
\subfigure[$\rho(r)$]{{\includegraphics[width=0.3\textwidth]{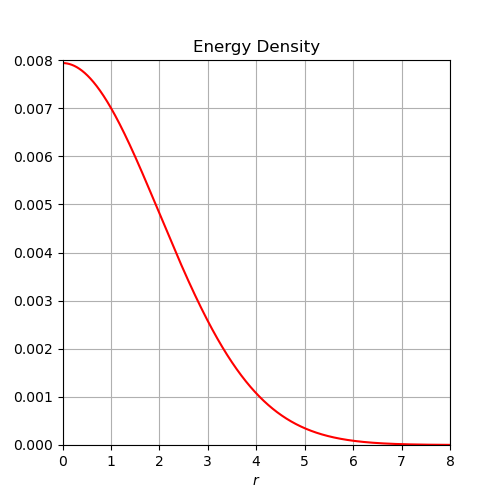}}
\label{fig: ED}}&
\subfigure[$\rho(r) + p_r(r)$]{{\includegraphics[width=0.3\textwidth]{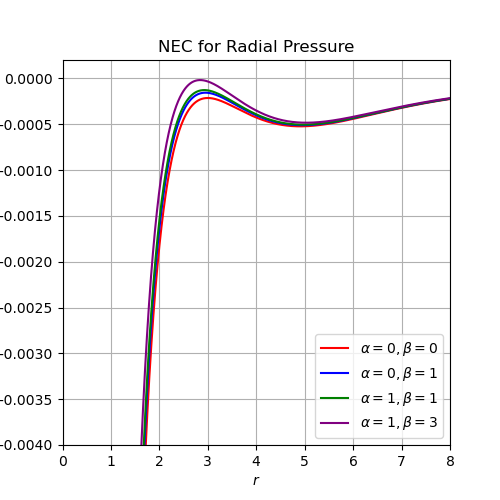}}
\label{fig: NECr}}&
\subfigure[$\rho(r) + p_t(r)$]{{\includegraphics[width=0.3\textwidth]{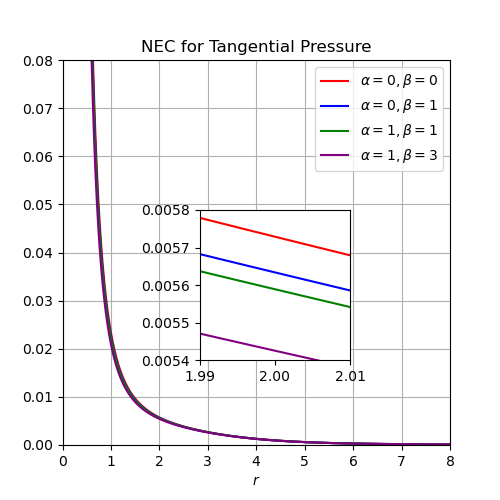}}
\label{fig: NECt}}
\end{tabular}
\caption{The figure illustrates the profile of energy density and NEC relative to the radial coordinate $r$ for different values of $\alpha$ and $\beta$ [showing GR $\left(\alpha=0, \beta=0\right)$, $f(R,T)$ $\left(\alpha=0, \beta=1\right)$, and $f(R,\mathcal{L}_m,T)$ $\left(\alpha=1, \beta\in\lbrace{1,3\rbrace}\right)$ cases] under the Gaussian distribution. The other parameters, specifically $\Theta = 2$, $M = 1$, and $r_0 = 1$, are held constant.}
\label{fig: EC}
\end{figure*}

Let us now employ this theoretical consideration for the Gaussian mass distribution. In Eqs. \eqref{15b} and \eqref{15c}, we substitute the Gaussian shape function, Eq. \eqref{b}, and obtain the pressure components
\begin{equation}\label{35}
    \hspace{-0.2cm}p_r = \frac{M}{4\pi r^3} \Biggl(\frac{re^{-\frac{r^2}{4 \Theta }}}{\sqrt{\pi\Theta}}- \text{erf}\left(\frac{r}{2 \sqrt{\Theta }}\right)- \widetilde{\mathcal{K}}_G\Biggr) ,
\end{equation}
\begin{multline}\label{36}
    \hspace{-0.2cm}p_t = \frac{M}{8\pi r^3}\Biggl(\text{erf}\left(\frac{r}{2 \sqrt{\Theta }}\right)-\frac{r(r^2+2\Theta)}{2\sqrt{\pi\Theta^{3}}}e^{-\frac{r^2}{4 \Theta }}
    +\widetilde{\mathcal{K}}_G\Biggr),
\end{multline}
where $\widetilde{\mathcal{K}}_G=\mathcal{K}_G+\frac{4\pi r_0}{\lambda M}$. At the throat, $r=r_0$, the Null Energy Conditions read
\begin{equation}
\begin{split}\label{eq: NECG}
\rho+p_r\bigg\vert_{r=r_0}&= \,\,\frac{M e^{-\frac{r_0^2}{4 \Theta }}}{(4\pi\Theta)^{3/2}}\,-\,\,\frac{1}{\lambda r_0^2 }\,\overset{!}{\geq}0\,,\\
\rho+p_t\bigg\vert_{r=r_0}&=\frac{M e^{-\frac{r_0^2}{4 \Theta }}}{2(4\pi\Theta)^{3/2}}+\frac{1}{2\lambda  r_0^2}\overset{!}{\geq}0\,.
\end{split}
\end{equation}
As expected, for the positive values of  $\beta+\frac{\alpha}{2}$ satisfying the FOC, the radial NEC condition is violated while the tangential one becomes trivial. This behavior is also visible in Figures \ref{fig: NECr} and \ref{fig: NECt}. However, restricting ourselves to negative $\beta+\frac{\alpha}{2}$ and again, taking $M=r_0=1$ and $\Theta=2$, Eq. \eqref{eq: NEC} demands $\beta+\frac{\alpha}{2}\lessapprox-168$. One can easily check that for this value, the inequalities of Eq. \eqref{eq: NECG} and the Flaring-Out Condition are satisfied. From Eq. \eqref{eq: NEC}, we find the general relation
\begin{equation}
    \beta+\frac{\alpha}{2}<-\frac{(4\pi\Theta)^{3/2}}{Mr_0^2}e^{\frac{r_0^2}{4\Theta}}-8\pi,
\end{equation}
for which non-exotic traversability is possible.
\begin{figure*}[t]
\begin{tabular}{ccc}
\subfigure[$\rho(r)$]{{\includegraphics[width=0.3\textwidth]{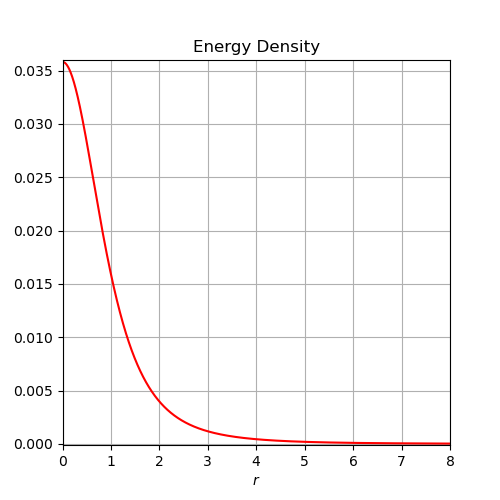}}
\label{fig: EDL}}&
\subfigure[$\rho(r) + p_r(r)$]{{\includegraphics[width=0.3\textwidth]{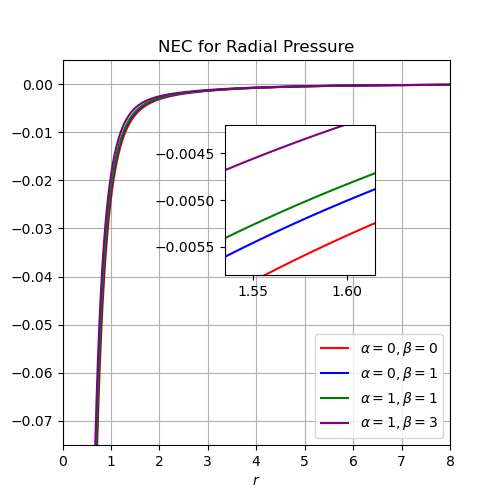}}
\label{fig: NECrL}}&
\subfigure[$\rho(r) + p_t(r)$]{{\includegraphics[width=0.3\textwidth]{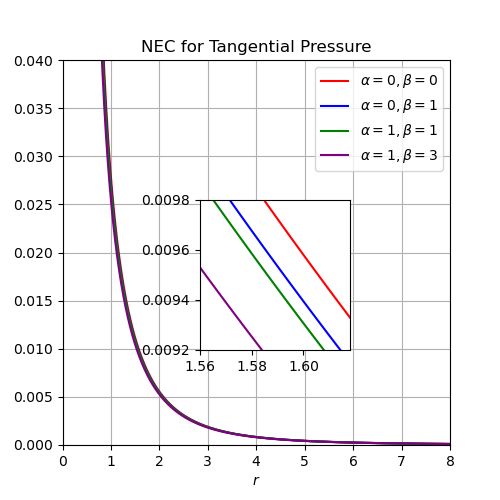}}
\label{fig: NECtL}}
\end{tabular}
\caption{The figure illustrates the profile of energy density and NEC relative to the radial coordinate $r$ for different values of $\alpha$ and $\beta$ [showing GR $\left(\alpha=0, \beta=0\right)$, $f(R,T)$ $\left(\alpha=0, \beta=1\right)$, and $f(R,\mathcal{L}_m,T)$ $\left(\alpha=1, \beta\in\lbrace{1,3\rbrace}\right)$ cases] under the Lorentzian distribution. The other parameters, specifically $\Theta = 2$, $M = 1$, and $r_0 = 1$, are held constant.}
\label{fig: ECL}
\end{figure*}

\subsection{Lorentzian Distribution}
Using the Lorentzian shape function, Eq. \eqref{34}, a similar calculation yields
\begin{multline}\label{37}
    \hspace{-0.2cm}p_r =
 \frac{M}{2\pi^2r^3} \Biggl(\frac{\sqrt\Theta r}{\Theta+r^2}-\tan ^{-1}\left(\frac{r}{\sqrt{\Theta }}\right)-\widetilde{\mathcal{K}}_L\Biggr)
\end{multline}
\begin{multline}\label{38}
    \hspace{-0.2cm}p_t = 
 \frac{M}{4\pi^2r^3} \Biggl(\tan ^{-1}\left(\frac{r}{\sqrt{\Theta }}\right)-\frac{\sqrt\Theta r(\Theta+3r^2)}{(\Theta+r^2)^2}+\widetilde{\mathcal{K}}_L \Biggr),
\end{multline}
where, $\widetilde{\mathcal{K}}_L=\mathcal{K}_L+\frac{2\pi^2r_0}{\lambda M}$. Again, for $r=r_0$, the NEC reduces to
\begin{equation}
\begin{split}\label{eq: NECL}
\rho+p_r\bigg\vert_{r=r_0}&=\,\,\frac{\sqrt{\Theta } M}{\pi ^2 \left(\Theta+r_0^2 \right)^2}\,-\,\,\frac{1}{\lambda r_0^2 }\,\overset{!}{\geq}0\,,\\
\rho+p_t\bigg\vert_{r=r_0}&=\frac{\sqrt{\Theta } M}{2 \pi ^2 \left(\Theta+r_0^2 \right)^2}+\frac{1}{2\lambda r_0^2}\overset{!}{\geq}0\,.
\end{split}
\end{equation}
Positive $\beta$ induces the contradiction between the FOC and the radial NEC previously explained, as one can see in Figures \ref{fig: NECrL}. Needless to say, the tangential NEC is trivially satisfied for these values of $\beta+\frac{\alpha}{2}$ (compare Figure \ref{fig: NECtL}). The restriction to negative $\beta+\frac{\alpha}{2}$ recovers the correct, non-exotic behavior; for the standard set of parameters, one must indeed have $\beta+\frac{\alpha}{2}\lessapprox-88$. Once again, this can be verified by inserting this value into Eqs. \eqref{eq: NECL} and the FOC, Eq. \eqref{eq: TWC}. The general formula obtained from the previous considerations yields:
\begin{equation*}
    \beta+\frac{\alpha}{2}<-\frac{\pi^2(\Theta+r_0^2)^2}{\sqrt{\Theta}Mr_0^2}-8\pi.
\end{equation*}
It is important to note that the model parameters, $\alpha$ and $\beta$, significantly influence all the energy conditions, particularly the NEC. It has been observed that when $-337.8<\alpha<53.9$, the NEC is violated near the throat $r_0=1$. Additionally, in situations where the NEC is violated, the contributions from $f(R,L_m,T)$ gravity are more than GR and $f(R,T)$ gravity.

\section{Gravitational Lensing Effects}
\label{ch: V}

\begin{figure*}[t]
  \includegraphics[width=0.3\textwidth]{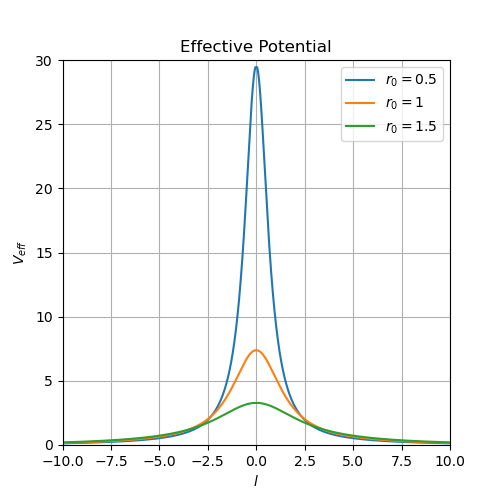}\,\,\,\,\,\,\,\,\,\,\,\,\,\,\,\,
  \includegraphics[width=0.3\textwidth]{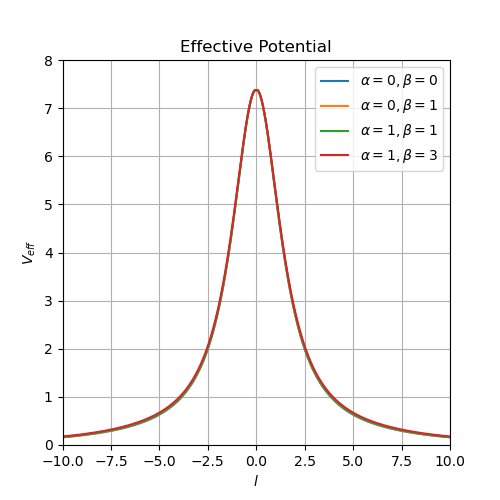}
\caption{The figure illustrates the profile of effective potential relative to the proper radial coordinate $l$ for different values of $r_0$, $\alpha$, and $\beta$ [showing GR $\left(\alpha=0, \beta=0\right)$, $f(R,T)$ $\left(\alpha=0, \beta=1\right)$, and $f(R,\mathcal{L}_m,T)$ $\left(\alpha=1, \beta\in\lbrace{1,3\rbrace}\right)$ cases] under the Gaussian distribution. The other parameters, specifically $\Theta = 2$, $M = 1$, and $r_0 = 1$, are held constant}
\label{fig: EPG}
\end{figure*}

\begin{figure*}[t]
  \includegraphics[width=0.3\textwidth]{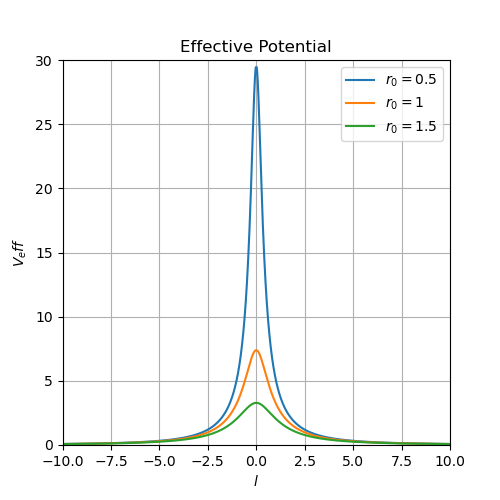}\,\,\,\,\,\,\,\,\,\,\,\,\,\,\,\,
  \includegraphics[width=0.3\textwidth]{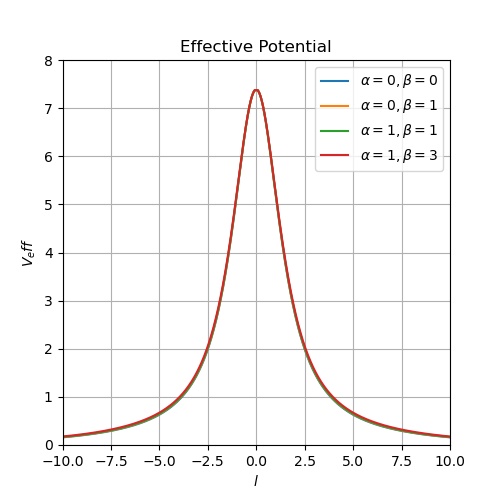}
\caption{The figure illustrates the profile of effective potential relative to the proper radial coordinate $l$ for different values of $r_0$, $\alpha$, and $\beta$ [showing GR $\left(\alpha=0, \beta=0\right)$, $f(R,T)$ $\left(\alpha=0, \beta=1\right)$, and $f(R,\mathcal{L}_m,T)$ $\left(\alpha=1, \beta\in\lbrace{1,3\rbrace}\right)$ cases] under the Lorentzian distribution. The other parameters, specifically $\Theta = 2$, $M = 1$, and $r_0 = 1$, are held constant.}
\label{fig: EPL}
\end{figure*}

This section will focus on the behavior of light-like particles in close proximity to the throat of the WH, the so-called \textit{gravitational lensing}. Various types of WHs, including charged \cite{V: GOD}, massless \cite{I: NAN}, Janis-Newman-Winnicour-WHs \cite{V: DEY} and Ellis-WHs \cite{V: TSU}, have been subject to contemporary research on lensing effects. Furthermore, lensing effects have also been used to examine the structure of black holes and naked singularities \cite{V: VIR, V: VIR1, V: BOZZA, V: BOZZA1}. In the following study, we employ a common approach found in the literature \cite{V: SHA, V: GOD}; for a comprehensive derivation of the formulas, we refer the reader to \cite{V: WEI}. We begin by stating the Lagrangian of a null curve $\gamma(s)=\bigl(t(s),r(s),\theta(s),\phi(s)\bigr)$ living in the spacetime given by Eq. (\ref{1}), with $\Dot{\gamma}=\partial\gamma/\partial s$:
\begin{equation}
    \mathcal{L}=-e^{-2\Phi}\frac{\Dot{t}^2}{2}+\frac{1}{1-b/r}\frac{\Dot{r}^2}{2}+r^2\frac{\Dot{\theta}^2}{2}+r^2\sin^2{\theta}\frac{\Dot{\phi}^2}{2}=0.
\end{equation}
On the one hand, spherical symmetry of $ds^2$ results in a vanishing $\Dot{\theta}$, enabling us to freely choose $\theta=\pi/2$. On the other hand, the translational and rotational symmetries of $t$ and $\phi$ allow us to rewrite the above equation using the conserved quantities $E:=\partial\mathcal{L}/\partial \Dot{t}=e^{-2\Phi}\Dot{t}$ and $L:=\partial\mathcal{L}/\partial \Dot{\phi}=r^2\Dot{\phi}$ as
\begin{equation}
    E^2=\frac{e^{2\Phi}}{1-b/r}\Dot{r}^2+V_{eff}(r),
\end{equation}
where $V_{eff}(r):=e^{2\Phi}L^2/r^2$. Identifying $E$ and $L$ with energy and angular momentum, respectively, the natural next step now would be to find the circular orbits for the light-like trajectories in the potential $V_{eff}$ (commonly dubbed \textit{photon spheres} \cite{V: VIR2}). Surprisingly, when searching for the minima of $V_{eff}$, it is easily seen that $\text{d}V_{eff}/\text{d}r\neq 0$ for a constant redshift - however, this is merely a result of a poor choice of coordinate system. Switching to proper, at $r_0$ non-singular coordinates $l(r)$ (where the two different values of $l$ corresponding to one $r$ can be interpreted as the two spacetime regions connected through the WH)
\begin{equation}
l(r)=\pm\bigintss_{\,r_0}^r\biggl(1-\frac{b(r')}{r'}\biggr)^{-\frac{1}{2}}\text{d}r',
\label{eq: l}
\end{equation}
the potential will have a maximum exactly at $r=r_0 (l=0)$ \cite{V: SHA}. Figures \ref{fig: EPG} and  \ref{fig: EPL} demonstrate this behavior for both the Gaussian and Lorentzian mass distributions for different values of $r_0$ and $\beta$ and the usual choice of additional parameters. Thus, the throat itself must serve as an unstable photon sphere. This fact can be further investigated by considering the angle $\bm{\alpha}$ by which the photon gets deflected:
\begin{equation}
    \bm{\alpha}(r_{tp})=-\pi+2\bigints_{\,r_{tp}}^\infty\frac{e^\Phi\bigl(1-\frac{b(r)}{r}\bigr)^{-\frac{1}{2}}}{\sqrt{\frac{r^2}{u^2}-e^{2\Phi}}}\text{d}r,
    \label{eq: DA}
\end{equation} where $u:=L/E$ is called the \textit{impact parameter}.
For constant redshift, $u$ is linearly related to the turning point $r_{tp}$ (characterized by $\Dot{r}(r_{tp})=0$):
\begin{equation}
\label{eq: u}
    u=r_{tp}e^{-\Phi},
\end{equation}
i.e., $\bm{\alpha}$ can also be considered a function of the ratio $L/E$. As $r=r_{0}$ is the sole photon sphere of this system, the only singularity in Eq. (\ref{eq: DA}) arises at $r_{tp}=r_0$ and for, $r_{tp}<r_0$, the function will diverge. The following sections will discuss the numerical results for different values of our parameters $\alpha$ and $\beta$.

\begin{figure*}[t]
\includegraphics[width=0.3\textwidth]{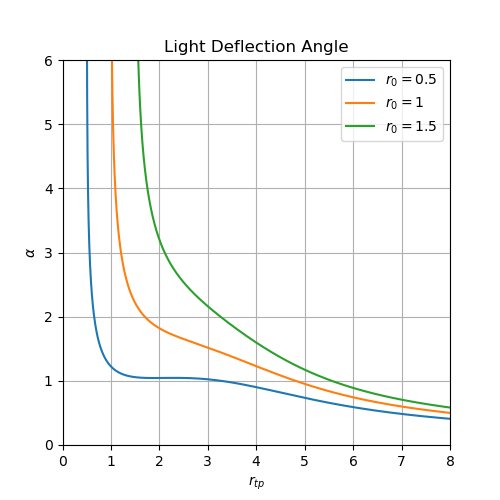}\,\,\,\,\,\,\,\,\,\,\,\,\,\,\,\,
\includegraphics[width=0.3\textwidth]{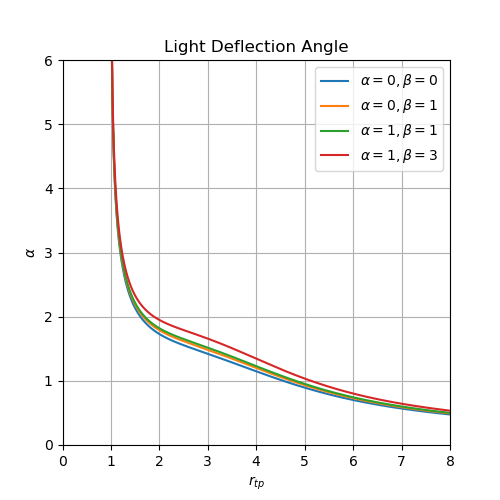}
\caption{The figure illustrates the profile of light deflection angle ($\bm{\alpha}$) relative to the radial coordinate $r_{tp}$ for different values of $r_0$, $\alpha$, and $\beta$ [showing GR $\left(\alpha=0, \beta=0\right)$, $f(R,T)$ $\left(\alpha=0, \beta=1\right)$, and $f(R,\mathcal{L}_m,T)$ $\left(\alpha=1, \beta\in\lbrace{1,3\rbrace}\right)$ cases] under the Gaussian distribution. The other parameters, specifically $\Theta = 2$, and $M = 1$, are held constant.}
\label{fig: LDAG}
\end{figure*}

\subsection{Gaussian Distribution}
Firstly, recall that in the case of $\alpha=1$ for $\beta\gtrapprox28$, $b/r>1$ for some values of $r>r_0$. Taking the asymptotic flatness into account, which holds for every $\alpha$ and $\beta$, this implies that $b/r=1$ for at least one $r>r_0$, resulting in a diverging deflection angle, Eq. \eqref{eq: DA} (implying that every light ray gets trapped in the WH). This behavior should be expected, as a violation of $b/r<1$ renders a diverging proper length coordinate, Eq. \eqref{eq: l}. Thus, we obtained a prime example of the physical meaning of non-traversability: for $b/r>1$, inside the WH, observers cannot physically exist, as there is no notion of proper length. Now, considering small values of positive $\beta+\frac{\alpha}{2}$, i.e. traversable WHs with couplings that require exotic matter, the results are in accordance with the literature \cite{I: NAN, V: GOD, V: DEY, V: SHA}; Figure \ref{fig: LDAG} demonstrates this for different values of $r_0$, $\alpha$ and $\beta$, respectively, with $\Phi=1$ and the remaining parameters set to their usual values. In every graph, the deflection angle diverges as $r_{tp}\rightarrow r_0$, and in the limit of large $r_{tp}$, there is no deflection (due to asymptotic flatness of the spacetime). On the contrary, the behavior drastically changes when restricting ourselves to $\beta+\frac{\alpha}{2}\lessapprox-168$, as suggested in Ch. \ref{ch: IV}. While both limits, $r_{tp}\rightarrow r_0$ and
$r_{tp}\rightarrow \infty$ remain unchanged (which should be demanded for consistency reasons), the deflection angle will be negative for all $r_{tp}$ larger than some $\Tilde{r}$. Hence, the WH acts repulsively in this domain. For instance, $r_0=1$ and $\beta+\frac{\alpha}{2}=-168$ (and the other parameters equal to the above investigations), we find negative deflections for all $r_{tp}\gtrapprox1.24$. Using Eq. \eqref{eq: u}, the energy of the light ray must therefore satisfy
\begin{equation*}
    E\gtrapprox2.2L
\end{equation*}
so it cannot be repulsed by the gravitational interaction with the WH. Interestingly, the transformation from attractive to repulsive interaction lies within the range of negative, non-exotic modifications, i.e., there are negative combinations of $\beta+\frac{\alpha}{2}>-168$ corresponding to both negative and positive deflection angles.

\begin{figure*}[t]
\includegraphics[width=0.3\textwidth]{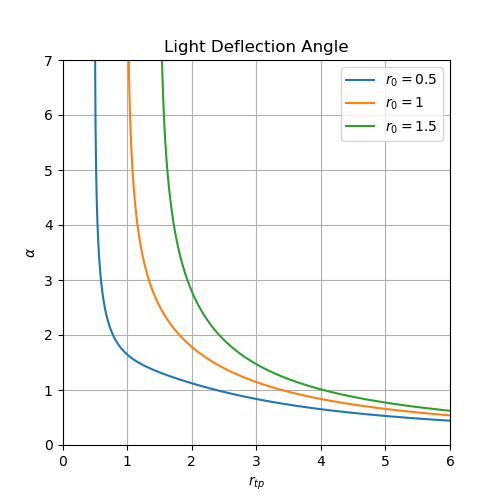}\,\,\,\,\,\,\,\,\,\,\,\,\,\,\,\,
\includegraphics[width=0.3\textwidth]{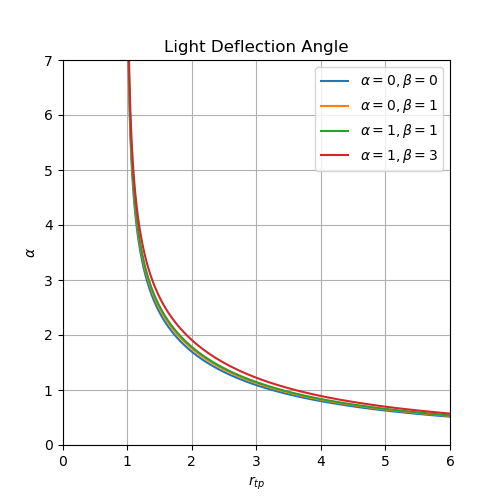}
\caption{The figure illustrates the profile of light deflection angle ($\bm{\alpha}$) relative to the radial coordinate $r_{tp}$ for different values of $r_0$, $\alpha$, and $\beta$ [showing GR $\left(\alpha=0, \beta=0\right)$, $f(R,T)$ $\left(\alpha=0, \beta=1\right)$, and $f(R,\mathcal{L}_m,T)$ $\left(\alpha=1, \beta\in\lbrace{1,3\rbrace}\right)$ cases] under the Lorentzian distribution. The other parameters, specifically $\Theta = 2$, and $M = 1$, are held constant.}
\label{fig: LDAL}
\end{figure*}

\subsection{Lorentzian Distribution}

The discussion of the Lorentzian mass distribution in the context of light ray deflection offers no new insights. Similar to the Gaussian case, here, for $r_0=M=\Phi=1$ and $\Theta=2$, $\alpha=1$ and $\beta\gtrapprox32$ result in the divergence of the deflection angle of every light ray and a cure in the form of a transformation to proper length coordinates is rendered impossible. Correct behavior, albeit demanding exotic mass to sustain the energy conditions, can be achieved for small positive couplings. Figure \ref{fig: LDAL} provides examples for $M=\Phi=1$ and $\Theta=2$ for different values of $r_0$, $\alpha$ and $\beta$, drawing a similar picture to the results of other researchers \cite{I: NAN, V: GOD, V: DEY, V: SHA}. The limits are physically justified, as described in the previous subsection. In the region of negative couplings that are not strong enough to support the WH without the existence of exotic matter, there will again be a certain value for which the gravitational interaction will turn into a repulsive force. Thus, light rays will generally be deflected away from the throat for $\beta+\frac{\alpha}{2}\lessapprox-88$ (the threshold for WHs supported by regular Lorentzian matter). Choosing $\beta+\frac{\alpha}{2}=-88$ and the other parameters as before, only for $r_{tp}\lessapprox1.48$ there is an attractive deflection, resulting in the condition 
\begin{equation*}
    E\gtrapprox1.8L\,
\end{equation*}
for reaching the throat of the WH (regarding light rays).

\section{Tolman-Oppenheimer-Volkoff Condition}\label{ch: VI}

The Tolman-Oppenheimer-Volkoff (TOV) equation \cite{Oppenheimer} describes the gravitational equilibrium for spherically symmetric spacetimes. It is commonly used to assess the stability of specific solutions in general relativity or modified gravity \cite{Gorini}. For an anisotropic mass distribution in regular GR, the TOV can be generalized to \cite{Kuhfittig}
\begin{eqnarray}\label{51}
\frac{\nu'}{2}(\rho+p_r)+\frac{dp_r}{dr}+\frac{2}{r}(p_r-p_t)=0,
\end{eqnarray}
 where, as we recall from Eq. \eqref{1}, $\nu=2\phi(r)$.\\
Identifying the terms with the hydrostatic, gravitational, and anisotropic force
\begin{equation}\label{52}
F_h=-\frac{dp_r}{dr}, ~~F_g=-\frac{\nu'}{2}(\rho+p_r), ~F_a=\frac{2}{r}(p_t-p_r),    
\end{equation}
equilibrium is achieved if $F_h+F_g+F_a=0$ holds. Our choice of constant redshift further simplifies this relation as $F_g$ vanishes:
\begin{equation}
\label{eq: eq}
    F_h+F_a=0.
\end{equation}
For our linearly modified gravity, the non-constancy of the energy-momentum tensor introduces the additional terms \cite{I: HAG}
\begin{equation}
    \nabla^\mu T_{\mu\nu}=-\frac{1}{\lambda}\Biggl[\Bigl(\frac{\alpha}{2}+\beta\Bigr)\nabla_\nu\rho+\frac{1}{2}\Bigl(\beta\nabla_\nu T-\frac{\alpha}{2}\nabla_\nu \rho\Bigr)\Biggr].
\end{equation}
However, the resulting force $F_m$ \cite{VI: DGU}
\begin{equation}
    F_m=\frac{\beta}{2\lambda}(\rho'+p'_r+2p'_t)
\end{equation}
vanishes, since $\rho+p_r+2p_t=0$ everywhere, compare Eq. \eqref{eq:SEC}.
Thus, from now on, Eq. \eqref{eq: eq} will serve as the condition for the stability of our solutions discussed in the following.

\begin{figure*}[t]
\begin{tabular}{cc}
\includegraphics[width=0.3\textwidth]{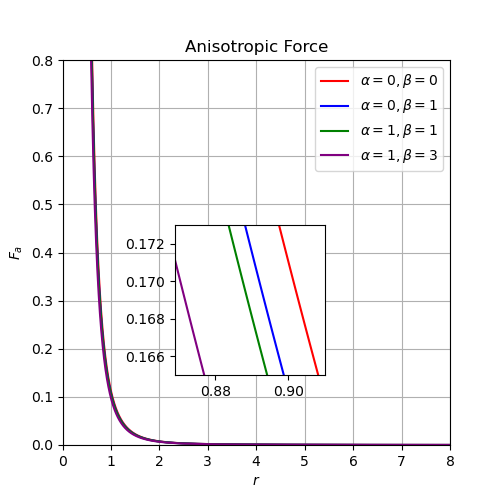}&
\includegraphics[width=0.3\textwidth]{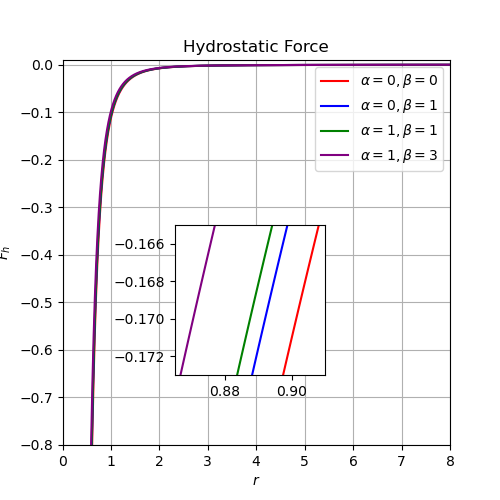}
\end{tabular}
\caption{The figure illustrates the changes in the anisotropic and hydrostatic forces relative to the radial coordinate $r$ for different values of $\alpha$ and $\beta$ [showing GR $\left(\alpha=0, \beta=0\right)$, $f(R,T)$ $\left(\alpha=0, \beta=1\right)$, and $f(R,\mathcal{L}_m,T)$ $\left(\alpha=1, \beta\in\lbrace{1,3\rbrace}\right)$ cases] under the Gaussian distribution. The other parameters, specifically $\Theta = 2$, $M = 1$, and $r_0 = 1$, are held constant.}
\label{fig: TOV}
\end{figure*}

\begin{figure*}[t]
\begin{tabular}{cc}
\includegraphics[width=0.3\textwidth]{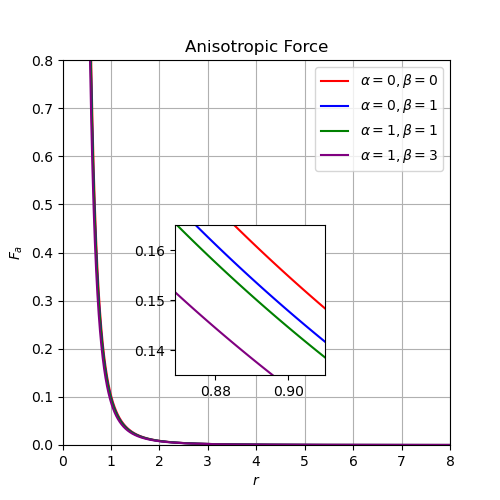}&
\includegraphics[width=0.3\textwidth]{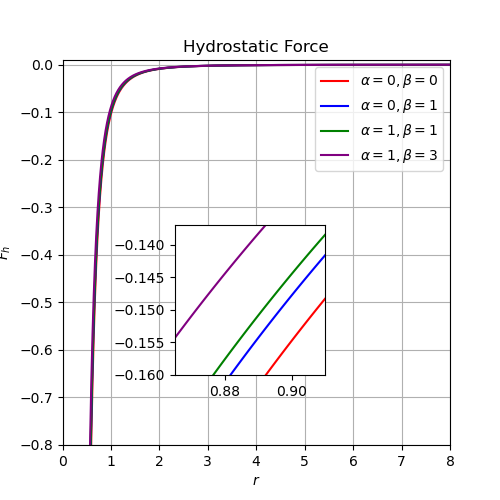}
\end{tabular}
\caption{The figure illustrates the changes in the anisotropic and hydrostatic forces relative to the radial coordinate $r$ for different values of $\alpha$ and $\beta$ [showing GR $\left(\alpha=0, \beta=0\right)$, $f(R,T)$ $\left(\alpha=0, \beta=1\right)$, and $f(R,\mathcal{L}_m,T)$ $\left(\alpha=1, \beta\in\lbrace{1,3\rbrace}\right)$ cases] under the Lorentzian distribution. The other parameters, specifically $\Theta = 2$, $M = 1$, and $r_0 = 1$, are held constant.}
\label{fig: TOVL}
\end{figure*}

\subsection{Gaussian Distribution}
Let us firstly calculate the respective forces, starting with the hydrostatic force by differentiating Eq. \eqref{35}:
\begin{equation}
        F_h = \left(1+\frac{6\Theta}{r^2}\right)\frac{Me^{-\frac{r^2}{4 \Theta}}}{r(4\pi\Theta)^{3/2}}
        - \frac{3M}{4\pi r^4}\left(\widetilde{\mathcal{K}}_G+\text{erf}\biggl(\frac{r}{2 \sqrt{\Theta }}\biggr)\right).
\end{equation}
Manipulating Eq. \eqref{35} and \eqref{36} according to the definition of the anisotropic force, Eq. \eqref{52}, yields
\begin{equation}
F_a=\frac{M}{4\pi r^4} \Biggl(3\,\text{erf}\left(\frac{r}{2 \sqrt{\Theta }}\right)-\frac{r(r^2+6\Theta)}{2\sqrt{\pi\Theta^3}}e^{-\frac{r^2}{4 \Theta }}+3\widetilde{\mathcal{K}}_G\Biggr).
\end{equation}
Carefully rearranging terms, we find that indeed $F_h+F_a=0$, i.e., the Gaussian WH is in gravitational equilibrium independently of the coupling strengths $\alpha$ and $\beta$. In Figure \ref{fig: TOV}, both forces are plotted with the common set of standard parameters, further emphasizing our result.

\subsection{Lorentzian Distribution}
Now, using Eqs. \eqref{37} and \eqref{38} analogously, we find: 

\begin{equation}
\begin{split}
        F_h = \left(3+\frac{2r^2}{\Theta+r^2}\right)\frac{M\sqrt
        \Theta}{2\pi^2r^3(\Theta+r^2)}\\
        -\frac{3M}{2\pi^2 r^4}\left(\widetilde{\mathcal{K}}_L+\tan
        ^{-1}\biggl(\frac{r}{\sqrt{\Theta }}\biggr)\right),
\end{split}
\end{equation}
\begin{equation}
F_a=\frac{M}{2\pi^2 r^4} \Biggl(3\tan^{-1}\left(\frac{r}{\sqrt{\Theta}}\right)-\frac{\sqrt{\Theta}r(3\Theta+5r^2)}{(\Theta+r^2)^2}+3\widetilde{\mathcal{K}}_L\Biggr).
\end{equation}
Here, too, a quick comparison tells us that the stability of the wormhole is verified via the TOV. The hydrostatic and anisotropic forces are visually represented in Fig. \ref{fig: TOVL} for $r_0=M=1$ and $\Theta=2$. Both forces are apparently of equal magnitude and different signs, as expected. 
\section{Conclusion}
\label{ch: VII}
In this study, we employed a linear $f(R,\mathcal{L}_m, T)$ model of modified gravity, which proved fruitful in the context of spherically symmetric wormholes with throat radius $r_0$ and mass $M$. Analytical solutions for the shape function $b(r)$ have been obtained from the field equation assuming a constant redshift function $\Phi$ and non-commutative (parametrized by $\Theta$) matter distributions $\rho$ of Gaussian and Lorentzian type. The Flaring-Out Condition, introduced by Morris and Thorne \cite{I: MOR} supplied us with upper bounds for the coupling parameters; 
\begin{equation}
        \beta+\frac{\alpha}{2}<\frac{(4\pi\Theta)^{3/2}}{Mr_0^2}e^{\frac{r_0^2}{4\Theta}}-8\pi
\end{equation} 
in the Gaussian case and 
\begin{equation}
        \beta+\frac{\alpha}{2}<\frac{\pi^2(\Theta+r_0^2)^2}{\sqrt{\Theta}Mr_0^2}-8\pi
\end{equation}
in the Lorentzian case. Here $\alpha$ and $\beta$ are the coupling strengths corresponding to the matter Lagrangian $\mathcal{L}_m$ and the trace term $T$, respectively. Furthermore, for $M=r_0=1$, $\Theta=2$ and choosing e.g. $\alpha=1$, $\beta\lessapprox28$ (Gaussian mass disbtribution) and $\beta\lessapprox32$ (Lorentzian mass distribution) s.t. $b/r<1$ everywhere. The stability of these solutions has been verified explicitly using the Tolman-Oppenheimer-Volkoff equation $F_g+F_h+F_a=0$, where we showed that while the $F_g$ vanishes due to the constant redshift, $F_h$ and $F_a$ will cancel independently of the coupling parameters $\alpha$ and $\beta$. Moreover, examining the generalized energy conditions, we found that at the throat of the wormhole, for
\begin{equation}
    \beta+\alpha/2\geq -8\pi,
\end{equation}
exotic matter cannot be avoided by our modification of gravity. However, choosing
\begin{equation}
    \beta+\alpha/2+8\pi<-\frac{1}{\rho r_0^2}
\end{equation}
the wormhole geometry can be supported by regular matter and the non-standard gravitational interaction. An immediate consequence of this choice of parameters is that the Weak, Strong, and Dominant Energy Conditions will be satisfied, too. However, during the discussion of gravitational lensing, we also learned that this also implied that, excluding the direct neighborhood of the throat, the gravitational force would act repulsively, resulting in lower bounds for the energy $E$ of the light rays that can reach the throat. For the Gaussian matter distribution and $=r_0=1$, $\Theta=2$ and $\beta+\frac{\alpha}{2}=-168$, the condition reads
\begin{equation}
    E\gtrapprox2.2L,
\end{equation}
where $L$ is the angular momentum of the photon. Choosing $\beta+\frac{\alpha}{2}=-88$ and the remaining parameters as before,
\begin{equation}
    E\gtrapprox1.8L,
\end{equation}
for the Lorentzian matter distribution. This result raises the question of whether there is a deeper reason for the fact that the circumvention of exotic matter demands couplings that imply repulsive gravitational interactions. 
Additionally, further research is also suggested regarding the modification itself; our model could serve as a first-order approximation of non-linear $f(R,\mathcal{L}_m, T)$, including the non-standard coupling of $R$, as well.\\
In \cite{VI: ZUBAIR1}, wormhole solutions were discussed with $f(R,T)=R+\lambda T$ model and obtained exact solutions in terms of exponential and hypergeometric functions under both distributions. Further, Godani \cite{VI: GODANI} and Shamir et al. \cite{VI: SHAMIR} examined the wormhole solutions in $f(R,T)$ gravity with both non-commutative distributions and showed the stability of wormhole using the TOV equation. Furthermore, Zubair et al. \cite{I: ZUBAIR} explored wormhole solutions in modified $f(R,T)$ gravity with linear and non-linear models under both Gaussian and Lorentzian distributions. Their analytical analysis of gravitational lensing revealed an infinite deflection angle at the wormhole throat. This paper discusses wormhole solutions under non-commutative geometry in the context of $f(R,\mathcal{L}_m, T)$ theory that has not been explored yet. Also, we have investigated the gravitational lensing numerically and noticed that the deflection angle diverges at the throat. In addition, we have discussed the generalized TOV equation with extra force due to matter-coupling gravity. However, this extra force does not affect the stability of the solutions. It is important to note that in this paper, zero tidal force, i.e., $\Phi'(r)=0$, was considered to study wormhole solutions. Therefore, studying wormholes with non-constant redshift functions in this $f(R,\mathcal{L}_m,T)$ gravity would be an interesting area of research.
\section*{Data Availability}
There are no new data associated with this article.

\acknowledgments NL acknowledges the German Academic Exchange Service (DAAD) for granting the RISE scholarship (Ref. No.: 57715964). MT acknowledges the University Grants Commission (UGC), New Delhi, India, for awarding the National Fellowship for Scheduled Caste Students (UGC-Ref. No.: 201610123801). PKS acknowledges National Board for Higher Mathematics (NBHM) under the Department of Atomic Energy (DAE), Govt. of India, for financial support to carry out the Research project No.: 02011/3/2022 NBHM(R.P.)/R\&D II/2152 Dt.14.02.2022. We warmly thank the honorable referee and the editor for the illuminating suggestions and constructive comments that have significantly improved our work in terms of research quality and presentation.

\end{document}